\documentclass[prd]{revtex4}
\usepackage{graphicx}
\usepackage{dcolumn}
\usepackage{bm}

\begin{document}

\title{Interactions between delta-like sources and potentials}

\author{G. T. Camilo}
 \email{gcamilo@unifei.edu.br}
\author{F. A. Barone}
 \email{fbarone@unifei.edu.br} 
\affiliation{ICE - Universidade Federal de Itajub\'a, Av. BPS 1303, Caixa Postal 50, CEP. 37500-903, Itajub\'a, MG, Brazil.}
\author{F. E. Barone}
 \email{febarone@cbpf.br} 
\affiliation{Laborat\'orio de F\'\i sica Experimental (LAFEX), Centro Brasileiro de Pesquisas F\'\i sicas, Rua Dr.\ Xavier Sigaud 150, Urca, CEP.  22290-180, Rio de Janeiro, RJ, Brazil.}

\date{\today}

%%%%%%%%
\begin{abstract}
The modified scalar boson propagator due to the presence of a hyperplane semi-transparent mirror is computed. From this,
the classical interaction between static charges and the mirror is investigated employing delta-like potentials and sources. 
Although the calculations for hyperplane mirrors are performed in arbitrary dimensions, and in a completely general way, it is shown that the results give rise to the usual image method as a particular case. The interaction between a point-like mirror and a point-like source is also considered in $3+1$ dimensions, where a central $1/R^{2}$ attractive potential is also obtained as a special case.
\end{abstract}
%%%%%%%%%

\maketitle

%%%%%%%%%
\section{Introduction}
%%%%%%%%%

Although it usually brings out notorious difficulties to deal with,  the realization of singularities has proven to be a requisite in the endeavor of understanding observable consequences of physical models in a multitude of different scenarios. Notably, singularities appear in classical physics, where the concept of point particles exhibits an outstanding usefulness, and in General Relativity where a singularity in the metric field is associated with the event horizon of black holes \cite{he}. Topological defects like domain walls, cosmic strings and monopoles are also related to singular field configurations \cite{defeitos}. 

Dirac delta function turns out to be the most suitable tool to cope with the realization of singular objects in any Field Theory. It provides a way out to the difficulties, modeling sources and potentials concentrated along arbitrary branes. As external sources it has been employed, for instance, to calculate the interaction energy intermediated by bosonic or even fermionic fields with different kinds of couplings in monopoles or multipoles distributions, and in arbitrary number of codimensions. In this way it is possible, from quantum principles, to recover and clarify novel forms of classical interactions and the topological mass generation mechanism, among other things \cite{Zee,BaroneHidalgo1,BaroneHidalgo2,BBH}. Closely related to this work, delta-like sources has also been used to investigate the interaction of the scalar and electromagnetic fields with two dimensional plasma sheets, in order to describe the pi-electrons of carbon nano-tubes \cite{Bordag2}.

On the other hand, delta-like potentials coupled to quantum fields have been used in many different contexts, mainly in the study of the Casimir effect \cite{Milton,Miltonlivro,Bordag,esferadelta,delta1d,Ricardo,Scardichio}, where the presence of physical objects, like semi-transparent mirrors, can be modeled by this kind of potential. Recently they have also been employed to unveil the physical aspects of the fermionic field in the so called soft MIT Bag Model \cite{softMit}.

A natural question that can be raised from the above discussion concerns in what kinds of modifications the scalar field propagator undergoes due to the presence of a single conductor, and its influence on static charges. This paper is devoted to an investigation in this guidance. All the results presented come solely from the classical field configuration in the presence of external sources and potentials. 

Specifically, we shall deal with a scalar field $\phi$, in $(D+1)-$dimensions and spacetime metric $\eta_{\mu\nu}=\textrm{diag }(+,-,-,\ldots,-)$. In section (\ref{hyperplane}) the interaction between a semi-transparent mirror and a point charge is investigated. The general result so obtained shows explicitly the dependence of the interaction energy on the distance between the mirror and the charge as well as on the mirror degree of transparency, and the classical image method is found as a particular case. In section (\ref{point}) the interaction energy between point-like charges and potentials is calculated in $(3+1)-$dimensions, from which an attractive potential with the same spatial dependence as the centrifugal one is also recovered as a particular case. Section (\ref{conclusion}) is devoted to final conclusions.

%%%%%%%%%
\section{Hyperplane Potential and Point Source}
\label{hyperplane}
%%%%%%%%%

In this section four-vectors shall be denoted by $x=(t,\mathbf{x})$, where $\mathbf{x}=(\mathbf{x_\parallel},x^{D})$, $\mathbf{x_\parallel}=(x^1,x^2,\ldots,x^{D-1})$ are the coordinates parallel to the hyperplane and $x^{D}$ is the perpendicular one. We also employ natural units where $\hbar=c=1$.

Here we shall consider a $(D-1)-$dimensional mirror lying along the hyperplane $x^D =0$. Its partial transparency shall be described by the potential $\frac{\mu}{2}\delta(x^{D})$, where $\mu>0$ is a coupling constant with appropriate dimension, establishing the degree of transparency of the hyperplane mirror. So, our starting point is the Lagrangian density
%%%%%%%%%%%%%%%%%%%%%%%%%%%%%%
\begin{equation}
\label{eq:1}
\mathcal L = \frac{1}{2}\partial_\nu\phi\partial^\nu\phi - \frac{1}{2}[m^2 + \mu\delta(x^{D})]\phi^2+J\phi\ .
\end{equation}
%%%%%%%%%

The presence of a point-like source is accomplished by the external current,
%%%%%%%%%
\begin{equation}
\label{defJ1}
J(x)=\lambda\delta^{(D)}(\mathbf{x}-\mathbf{a})\ ,
\end{equation}
%%%%%%%%%
where $\lambda$ is the coupling constant and $\mathbf{a}$ is a constant vector standing for the charge position.

Here the Green function of the theory, $G(x,x')$, is defined as the particular solution of the equation, 
%%%%%%%%%%%%%%%%%%%%%%%%%%%%%%
\begin{equation}
\label{eq:3}
[\partial_\nu\partial^\nu+m^2 + \mu\delta(x^{D})]G(x,x') = \delta^{(D+1)}(x-x')\ .
\end{equation}
%%%%%%%%%%%%%%%%%%%%%%%%%%%%%%

It is worth mentioning that the Green function in (\ref{eq:3}) is defined with the implicit prescription of a small negative imaginary part for the mass, $m\to m-i\epsilon,\ \epsilon>0$, inserted in order to ensure convergence of functional integrals \cite{DasPI}. If we take $\mu=0$, $G(x,x')$ reduces to the Feynman Green function for the Klein-Gordon field. In this sense, the particular solution of (\ref{eq:3}) can be taken as a generalization of the Feynman Green function for the Klein-Gordon field in the presence of a delta potential. All over this paper we shall omit this implicit imaginary part for the mass just for convenience.

The interaction energy of the system can be found according to the expression,
%%%%%%%%%%%%%%%%%%%%%%%%%%%%%%
\begin{equation}
\label{eq:5}
E=\lim_{T\rightarrow\infty}-\frac{1}{2T}\int d^{D+1}x\int d^{D+1}x'J(x)G(x,x')J(x')\ .
\end{equation}
%%%%%%%%%%%%%%%%%%%%%%%%%%%%%%

Once the results we shall calculate in this section are obtained directly from expression (\ref{eq:5}), they correspond to classical interactions.

Substituting the source (\ref{defJ1}) in the above expression and integrating the delta functions we have,
%%%%%%%%%%%%%%%%%%%%%%%%%%%%%%
\begin{equation}
\label{eq:E}
E=\lim_{T\rightarrow\infty}-\frac{\lambda^2}{2T}\int_{-T/2}^{T/2} dt\int_{-T/2}^{T/2}dt'G(t,\mathbf{a};t',\mathbf{a}) \ .
\end{equation}
%%%%%%%%%%%%%%%%%%%%%%%%%%%%%%	

At this point, our task is to solve equation (\ref{eq:3}) for the Green function $G$. For convenience, we write $\partial_\nu\partial^\nu=\partial_t^2-\partial_{\mathbf{x_\parallel}}^2-\partial_{x^{D}}^2$ and Fourier transform $G$ in the first $D$ spacetime coordinates ($t,\mathbf{x}_\parallel$) as follows,
%%%%%%%%%%%%%%%%%%%%%%%%%%%%%% 
\begin{eqnarray}
\label{eq:6}
G(x,x')=\int\frac{d\omega}{2\pi}e^{-i\omega(t-t')}\int\frac{d^{D-1}\mathbf{k_\parallel}}{(2\pi)^{D-1}}e^{i\mathbf{k_\parallel}\cdot(\mathbf{x_\parallel}-
\mathbf{x_\parallel}')}
\ {\cal G}(\omega,\mathbf{k_\parallel};x^{D}\!,{x'}^{D}) \ .
\end{eqnarray}
%%%%%%%%%%%%%%%%%%%%%%%%%%%%%%

Substituting (\ref{eq:6}) into (\ref{eq:3}), it is straightforward to see that the reduced Green's function ${\cal G}(\omega,\mathbf{k_\parallel};x^{D},x'^{D})$ must be the particular solution of the equation
%%%%%%%%%%%%%%%%%%%%%%%%%%%%%%
\begin{eqnarray}
\label{eq:7}
[-\partial_{x^{D}}^2\! + \sigma^2\! + \mu\delta(x^{D})]{\cal G}(\omega,\mathbf{k_\parallel};x^{D}\!,{x'}^{D}) = \delta(x^{D}\!\!\!-{x'}^{D}) \ ,
\end{eqnarray}
%%%%%%%%%%%%%%%%%%%%%%%%%%%%%%
where we have defined $\sigma^2\equiv\mathbf{k_\parallel}^2+m^2-\omega^2$ (with an implicit small negative imaginary part for the mass).
 
The solution for (\ref{eq:7}) can be easily checked to be given recursively in integral form as, 
%%%%%%%%%%%%%%%%%%%%%%%%%%%%%%
\begin{eqnarray}
\label{eq:8}
{\cal G}(\omega,\mathbf{k_\parallel};x^{D}\!,{x'}^{D})&=&{\cal G}_0(\omega,\mathbf{k_\parallel};x^{D}\!,{x'}^{D})
- \int dy {\cal G}(\omega,\mathbf{k_\parallel};x^{D}\!,y)\mu\delta(y)
{\cal G}_{0}(\omega,\mathbf{k_\parallel};y,{x'}^{D})\cr\cr
&=& {\cal G}_0(\omega,\mathbf{k_\parallel};x^{D}\!,{x'}^{D})
- \mu {\cal G}(\omega,\mathbf{k_\parallel};x^{D}\!,0){\cal G}_{0}(\omega,\mathbf{k_\parallel};0,{x'}^{D})\ ,%\cr
\end{eqnarray}
%%%%%%%%%%%%%%%%%%%%%%%%%%%%%%
where ${\cal G}_0(\omega,\mathbf{k_\parallel};x^{D},{x'}^{D})$ solves the corresponding equation with no potential
%%%%%%%%%%%%%%%%%%%%%%%%%%%%%%
\begin{equation}
\label{eq:9}
[-\partial_{x^{D}}^2 + \sigma^2]{\cal G}_0(\omega,\mathbf{k_\parallel};x^{D},{x'}^{D}) = \delta(x^{D}-{x'}^{D})\ .
\end{equation}
%%%%%%%%%%%%%%%%%%%%%%%%%%%%%%
In the notation of reference \cite{Miltonlivro}, ${\cal G}_0(\omega,\mathbf{k_\parallel};x^{D},{x'}^{D})$ is the reduced Feynman Green's function of the Klein-Gordon field.

An alternative procedure to justify eq. (\ref{eq:8}) consists in considering the fact that the solution for (\ref{eq:3}) is given by a kind of Bethe-Salpeter equation
%%%%%%%%%
\begin{equation}
G(x,x')=G_{0}(x,x')-\int d^{D+1}y\ G(x,y)\mu\delta(y^{D})G_{0}(y,x')\ ,
\end{equation}
%%%%%%%%%
where $G_{0}(x,x')$ is the Klein-Gordon Green's function, which solves (\ref{eq:3}) without potential. With the Fourier transform (\ref{eq:6}) and a similar expression for $G_{0}(x,x')$, whose reduced Green's function is ${\cal G}_0(\omega,\mathbf{k_\parallel};x^{D},{x'}^{D})$, we are taken to (\ref{eq:8}).

By setting ${x'}^{D}=0$ in the second line of (\ref{eq:8}), and performing some simple manipulations, we can obtain ${\cal G}(\omega,\mathbf{k_\parallel};x^{D},0)$ strictly in terms of ${\cal G}_{0}(\omega,\mathbf{k_\parallel};x^{D},0)$. Using the result back again in the second line of (\ref{eq:8}) we are taken to,
%%%%%%%%%%%%%%%%%%%%%%%%%%%%%%
\begin{eqnarray}
\label{eq:10}
{\cal G}(\omega,\mathbf{k_\parallel};x^{D},{x'}^{D}) = {\cal G}_0(\omega,\mathbf{k_\parallel};x^{D},{x'}^{D}) - \frac{\mu {\cal G}_0(\omega,\mathbf{k_\parallel};x^{D},0){\cal G}_0(\omega,\mathbf{k_\parallel};0,{x'}^{D})}{1+\mu {\cal G}_0(\omega,\mathbf{k_\parallel};0,0)} \qquad .
\end{eqnarray}
%%%%%%%%%%%%%%%%%%%%%%%%%%%%%%

At this point we need to calculate ${\cal G}_0(\omega,\mathbf{k_\parallel};x^{D},{x'}^{D})$, which means to solve equation (\ref{eq:9}). It can be done as usual, by the Fourier method \cite{Bordag,Ricardo}, and the result is 
%%%%%%%%%%%%%%%%%%%%%%%%%%%%%%
\begin{equation}
\label{eq:11}
{\cal G}_0(\omega,\mathbf{k_\parallel};x^{D},{x'}^{D}) = \int\frac{dk^D}{2\pi}\frac{e^{ik^D(x^{D}-{x'}^{D})}}{(k^D)^2+\sigma^2} = \frac{e^{-\sigma |x^{D}-{x'}^{D}|}}{2\sigma} \ .
\end{equation}
%%%%%%%%%%%%%%%%%%%%%%%%%%%%%%	

Now we point out that the first term ${\cal G}_0(\omega,\mathbf{k_\parallel};z,z')$ in equation (\ref{eq:10}) will not contribute to the interaction energy between the charge and the hyperplane. The contribution which comes from ${\cal G}_0$ is obviously associated with the self interaction of the source.  As a matter of fact, ${\cal G}_{0}$ itself is associated with the energy of the scalar field without the presence of the plane and its contribution to the energy does not depend on the distance between the brane and the charge. Hence it does not affect the force between them, and from now on we shall discard it. 
	
By taking into account equations (\ref{eq:11}), (\ref{eq:10}) and (\ref{eq:6}) and using a coordinate system where $\mathbf{a}=(0,\ldots,0,a)$ (there is no loss of generality in doing this) we can write the interaction energy (\ref{eq:E}) as
%%%%%%%%%
\begin{eqnarray}
E_{int} = \lim_{T\rightarrow\infty}\frac{-\lambda^2}{2T}\int_{-T/2}^{T/2} dt\int_{-T/2}^{T/2} dt' \int_{-\infty}^{\infty}\frac{d\omega}{2\pi} e^{-i\omega(t-t')}
\int\frac{d^{D-1}\mathbf{k_\parallel}}{(2\pi)^{D-1}}\left[-\frac{\mu e^{-2\sigma|a|}}{2\sigma(2\sigma+\mu)} \right]\ . 	
\end{eqnarray}
%%%%%%%%%	
It is worth mentioning that $|a|$ appearing above is the distance between the plane and the point-like source.

With the aid of the Fourier representation of Dirac delta functions $\delta(\omega)=\int dt/2\pi\ \exp(-i\omega t)$ and identifying the time interval as $T=\int_{-T/2}^{T/2}dt$, we have
%%%%%%%%%%%%%%%%%%%%%%%%%%%%%%
\begin{eqnarray}
E_{int}
=\frac{\mu\lambda^2}{2(2\pi)^{D-1}}\int d^{D-1}\mathbf{k_\parallel}
\frac{e^{-2|a|\sqrt{\mathbf{k_\parallel}^2+m^2}}}{2\sqrt{\mathbf{k_\parallel}^2+m^2}\left(2\sqrt{\mathbf{k_\parallel}^2+m^2}+\mu\right)}\ .
\end{eqnarray}
%%%%%%%%%%%%%%%%%%%%%%%%%%%%%%	
 	
The integral above can be simplified if we change from cartesian to hyperspherical coordinates, as discussed in references \cite{BaroneHidalgo1,Kaku}. The result is
%%%%%%%%%%%%%%%%%%%%%%%%%%%%%%
\begin{equation}
\label{eq:14}
E_{int} = \frac{\mu\lambda^2}{2(2\pi)^{D-1}}\ \  \Omega \int_0^\infty dr \frac{r^{D-2}e^{-2|a|\sqrt{r^2+m^2}}}{2\sqrt{r^2+m^2}\left(2\sqrt{r^2+m^2}+\mu\right)} \ ,
\end{equation}
%%%%%%%%%%%%%%%%%%%%%%%%%%%%%% 
where we have defined the total solid angle of the $(D-1)-$sphere as $\Omega \equiv 2\pi^{(D-1)/2}\Gamma\bigl[(D-1)/2\bigr]$, with $\Gamma(x)$ being the Gamma function \cite{Arfken}.

Performing the change of variable $r\to y=2\sqrt{r^2+m^2}$ in (\ref{eq:14}) we obtain the following expression,
%%%%%%%%%%%%%%%%%%%%%%%%%%%%%%
\begin{eqnarray}
\label{eq:Eint}
E_{int}(a,m,\mu,D) = \frac{\mu\lambda^2}{4(4\pi)^{(D-1)/2}\Gamma((D-1)/2)}
 \int_{2m}^\infty dy \left[\frac{y^2}{4}-m^2\right]^{\frac{D-3}{2}} \frac{e^{-|a|y}}{y+\mu}\ .
\end{eqnarray}
%%%%%%%%%%%%%%%%%%%%%%%%%%%%%% 
This is the general result for the interaction energy between a partially transparent mirror and a charge, but unfortunately this integral cannot be solved analytically for arbitrary values of $m$, $D$ and $\mu$. In order to have a better insight on the meaning of expression (\ref{eq:Eint}), let us analyze some special cases.

The first case of interest is the massless one, which is analytically solvable for arbitrary $\mu,D$. By putting $m=0$ in (\ref{eq:Eint}) the energy becomes
%%%%%%%%%%%%%%%%%%%%%%%%%%%%%%
\begin{eqnarray}
\label{eq:m=0}
E_{int}(a,m=0,\mu,D)
 = \frac{\mu\lambda^2}{(16\pi)^{(D-1)/2}\Gamma((D-1)/2)}\int_{0}^\infty dy\ \frac{y^{D-3}}{y+\mu}\ e^{-|a|y}\ .
\end{eqnarray}
%%%%%%%%%%%%%%%%%%%%%%%%%%%%%%
With the aid of the identity
%%%%%%%%%
\begin{equation}
y^{D-3}\ e^{-|a|y} = (-1)^{D-3}\frac{d^{D-3}}{d(|a|)^{D-3}}\ e^{-|a|y}\ ,
\end{equation}
%%%%%%%%%
also valid with the definition $d^{-1}/d(|a|)^{-1}=\int_{\infty}^{|a|}d|a|$, and performing the substitution $z=(y+\mu)/\mu$, the energy (\ref{eq:m=0}) becomes 
%%%%%%%%%%%%%%%%%%%%%%%%%%%%%%
\begin{eqnarray}
\label{eq:m=03}
E_{int}(a,m=0,\mu,D) = \frac{(-1)^{D-1}\mu\lambda^2}{(16\pi)^{(D-1)/2}\Gamma((D-1)/2)}
\frac{d^{D-3}}{d(|a|)^{D-3}}\left\{
Ei(1,|a|\mu)\ e^{|a|\mu}\right\} \ ,
\end{eqnarray}
%%%%%%%%%%%%%%%%%%%%%%%%%%%%%%    
where $Ei(u,v)$ is the exponential integral function \cite{Arfken}.

Another particular case of interest is the limit $\mu\rightarrow\infty$, corresponding physically to a field subjected to Dirichlet boundary conditions at the plane. In this limit, equation (\ref{eq:Eint}) reads 
%%%%%%%%%%%%%%%%%%%%%%%%%%%%%%
\begin{eqnarray}
\label{eq:mu}
E_{int}(a,m,\mu\to\infty,D) = \frac{\lambda^2}{4(4\pi)^{(D-1)/2}\Gamma((D-1)/2)}\int_{2m}^\infty dy
\left[\frac{y^2}{4}-m^2\right]^{(D-3)/2}\ e^{-|a|y}
\end{eqnarray}
%%%%%%%%%%%%%%%%%%%%%%%%%%%%%% 	 
which can be solved \cite{Gradstein} for general $m$ and $D$ to give
%%%%%%%%%%%%%%%%%%%%%%%%%%%%%%
\begin{eqnarray}
\label{eq:mum}
E_{int}(a,m,\mu\to\infty,D)
 = \frac{\lambda^2 m^{D-2}}{2(2\pi)^{D/2}}(2m|a|)^{1-(D/2)}K_{(D/2)-1}(2m|a|)\ ,
\end{eqnarray}
%%%%%%%%%%%%%%%%%%%%%%%%%%%%%%
where $K_\nu(x)$ is the modified Bessel function \cite{Arfken}. As expected for the scalar field \cite{Zee,BaroneHidalgo1}, in comparison with the electromagnetic one, we have an overall minus sign. 

Expression (\ref{eq:mum}) is the Yukawa-like interaction between the charge and its image related to the $(D-1)$-dimensional mirror in arbitrary codimensions. This fact can be verified by comparison with the results obtained in reference \cite{BaroneHidalgo1}, taking into account that, in (\ref{eq:mum}), the distance between the source and its image is $2|a|$. 

The massless case can be obtained from (\ref{eq:mum}) as exposed in \cite{BaroneHidalgo1}. The result is the Coulomb interaction in $D+1$ dimensions between the point charge and its image. 

\begin{figure}[!h]
 \centering
   \includegraphics[scale=0.20]{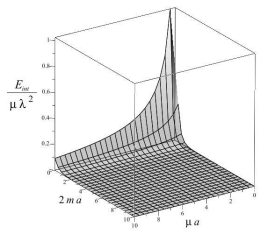}
  \caption{$E_{int}/\mu\lambda^{2}$ for $D=3$.}
  \label{fig1}
\end{figure}

From now on let us restrict ourselves to the $3+1$ space-time, corresponding to $D=3$. In this case the energy (\ref{eq:Eint}) can be easily integrated,
%%%%%%%%%%%%%%%%%%%%%%%%%%%%%%
\begin{equation}
\label{eq:D=3}
E_{int}(a,m,\mu,D=3) = \frac{\mu\lambda^2 }{16\pi}e^{|a|\mu} Ei(1,2m|a|+|a|\mu)\ .
\end{equation}
%%%%%%%%%%%%%%%%%%%%%%%%%%%%%% 
Result (\ref{eq:D=3}) is the generalization of the image method for a semi-transparent mirror with arbitrary mass for the field. The interaction energy falls off very quickly as the separation distance $a$ increases. This fall is determined by the field mass $m$ as well as the coupling parameter $\mu$, both of which acting in such a way to decrease the interaction energy between the mirror and the charge, as one can see from Figure (\ref{fig1}). 
%%%%%%%%%%%%%%%%%%%%%%%%%%%%
\\
%%%%%%%%%%%%%%%%%%%%%%%%%%%%

An interesting fact turns out when we consider the limit $\mu\rightarrow\infty$ in Eq. (\ref{eq:D=3}). As it has already been stated above, in this limit the plane corresponds to a perfect $2-$dimensional mirror, characterized by the Dirichlet's boundary condition along it. So, by taking $\mu\to\infty$ in expression (\ref{eq:D=3}) we obtain
%%%%%%%%%%%%%%%%%%%%%%%%%%%%%%
\begin{equation}
\label{eq:img}
E_{int}(a,m,\mu\to\infty,D=3) = \frac{\lambda^2 }{8\pi} \frac{e^{-2|a|m}}{2|a|}
\end{equation}
%%%%%%%%%%%%%%%%%%%%%%%%%%%%%% 
which is the 3-dimensional Yukawa potential between two charges at a distance $2|a|$ apart. If one prefers a massless field, it is enough to set $m=0$ to get the corresponding Coulomb potential with an overall minus sign. In both cases, the energy (\ref{eq:img}) is equivalent to the classical static potential calculated via image method.

%%%%%%%%%
\section{Point-like Potential and Source}
\label{point}
%%%%%%%%%

In this section we consider a $(3+1)-$dimensional spacetime with a point-like potential, $V({\bf x})=\mu\delta^{3}({\bf x}-{\bf b})$, interacting with a point-like external source $J({\bf x})=\lambda\delta^{3}({\bf x}-{\bf a})$. The potential and the source are concentrated at the points $\bf b$ and $\bf a$, respectively. Both the potential and the external current are taken to be stationary, meaning that ${\bf a}$ and ${\bf b}$ are constant vectors.
The corresponding lagrangian reads
%%%%%%%%%
\begin{equation}
\label{segundomodelo}
{\cal L}=\frac{1}{2}\partial_{\mu}\phi\partial^{\mu}\phi-\frac{1}{2}[m^{2}+V({\bf x})]\phi^{2}+J({\bf x})\phi\ .
\end{equation}
%%%%%%%%%
As discussed in reference \cite{Scardichio}, in order to avoid tachyonic field states, we must have the condition $m>4\pi/\mu$ if $\mu>0$. If $\mu<0$ the mass $m$ can have any (non-negative) value. 

In order to calculate the interaction energy between the potential and the source, we start again with the general expression (\ref{eq:E}), but this time it reads,
%%%%%%%%%
\begin{equation}
\label{energiageral}
E=\lim_{T\rightarrow\infty}-\frac{1}{2T}\int d^{4}xd^{4}yJ(x)G^{P}(x,y)J(y)\ ,
\end{equation}
%%%%%%%%%
where $E$ is the system energy and $G^{P}(x,y)$ the Green function of the model (\ref{segundomodelo}). The superscript $P$ stands for point-like potential. 

Once $G^{P}(x,y)$ is associated with the Klein-Gordon operator with the presence of the point-like potential, we have (as in the previous section, we are omitting a small negative imaginary part for the mass $m$).
%%%%%%%%%
\begin{equation}
[\partial_{\mu}\partial^{\mu}+m^{2}+V({\bf x})]G^{P}(x,y)=\delta^{4}(x,y)\ .
\end{equation}
%%%%%%%%%

The first task is to get $G^{P}(x,y)$, what can be done with the aid of the relation
%%%%%%%%%
\begin{eqnarray}
\label{zxc1}
G^{P}(x,y)&=&G_{0}(x,y)+\Delta G(x,y)\ ,
\end{eqnarray}
%%%%%%%%%
where we have defined
%%%%%%%%%
\begin{equation}
\label{defDeltaG}
\Delta G(x,y)=-\int d^{4}z G^{P}(x,z)V(z)G_{0}(z,y)
\end{equation}
%%%%%%%%%
and $G_{0}(x,y)$ is the free Klein-Gordon Feynman Green's function in the sense that
%%%%%%%%%
\begin{equation}
[\partial_{\mu}\partial^{\mu}+m^{2}]G_{0}(x,y)=\delta^{4}(x,y).
\end{equation}
%%%%%%%%%

Substituting (\ref{zxc1}) in (\ref{energiageral}) and neglecting the contribution which comes from the first term on the right hand side of (\ref{zxc1}), which is related to the self energy of the source, we have
%%%%%%%%%
\begin{equation} 
\label{zxc2}
E_{int}=-\frac{1}{2T}\int d^{4}xd^{4}yJ(x)\Delta G(x,y)J(y)\ .
\end{equation}
%%%%%%%%%

Up to now our analysis was quite general. From now on let us specify for the potential $V({\bf x})=\mu\delta^{3}({\bf x}-{\bf b})$. First we write the Green's functions $G^{P}$ and $G_{0}$ as Fourier integrals in the time variable, in such a way that
%%%%%%%%%
\begin{eqnarray}
\label{zxc3}
G^{P}(x,y)&=&\int\frac{d\omega}{2\pi}{\tilde{\cal G}}(\omega;{\bf x},{\bf y})e^{-i\omega(x^{0}-y^{0})}\ ,\cr\cr
G_{0}(x,y)&=&\int\frac{d\omega}{2\pi}{\tilde{\cal G}}_{0}(\omega;{\bf x},{\bf y})e^{-i\omega(x^{0}-y^{0})}\ .
\end{eqnarray}
%%%%%%%%%
So equation (\ref{zxc1}) reads
%%%%%%%%%
\begin{equation}
\label{zxc4}
{\tilde{\cal G}}(\omega;{\bf x},{\bf y})={\tilde{\cal G}}_{0}(\omega;{\bf x},{\bf y})
-\mu{\tilde{\cal G}}(\omega;{\bf x},{\bf b}){\tilde{\cal G}}(\omega;{\bf b},{\bf y})\ ,
\end{equation}
%%%%%%%%%
where we have used the explicit form of $V({\bf x})$, the definition (\ref{defDeltaG}) and performed the integral in $d^{4}z$.

Taking ${\bf y}={\bf b}$ in (\ref{zxc4}) and making some simple manipulations we can write ${\tilde{\cal G}}(\omega;{\bf x},{\bf b})$ in terms of ${\tilde{\cal G}}_{0}(\omega;{\bf x},{\bf b})$ and ${\tilde{\cal G}}_{0}(\omega;{\bf b},{\bf b})$. Putting the result back into (\ref{zxc4}) we are taken to
%%%%%%%%%
\begin{equation}
{\tilde{\cal G}}(\omega;{\bf x},{\bf y})={\tilde{\cal G}}_{0}(\omega;{\bf x},{\bf y})
-\frac{\mu{\tilde{\cal G}}_{0}(\omega;{\bf x},{\bf b}){\tilde{\cal G}}_{0}(\omega;{\bf b},{\bf y})}{1+\mu {\tilde{\cal G}}_{0}(\omega;{\bf b},{\bf b})}\ .
\end{equation}
%%%%%%%%%

Performing the Fourier integral in the above expression, as in Eq. (\ref{zxc3}), and comparing the result with (\ref{zxc1}) we can write
%%%%%%%%%
\begin{equation}
\Delta G(x,y)=-\int\frac{d\omega}{2\pi}e^{-i\omega(x^{0}-y^{0})}\frac{\mu{\tilde{\cal G}}_{0}(\omega;{\bf x},{\bf b}){\tilde{\cal G}}_{0}(\omega;{\bf b},{\bf y})}{1+\mu {\tilde{\cal G}}_{0}(\omega;{\bf b},{\bf b})}\ ,
\end{equation}
%%%%%%%%%
so the interaction energy (\ref{zxc2}) becomes
%%%%%%%%%
\begin{equation}
\label{zxc6}
E_{int}=\frac{\lambda^{2}\mu}{2}\frac{\Bigl({\tilde{\cal G}}_{0}(0;{\bf b},{\bf a})\Bigr)^{2}}{1+\mu {\tilde{\cal G}}_{0}(0;{\bf b},{\bf b})}\ ,
\end{equation}
%%%%%%%%%
where we have integrated over $d{\bf x}^{3}$, $d{\bf y}^{3}$, $dx^{0}$, $d\omega$ and $dy^{0}$ (in this order), used the Fourier representation $2\pi\delta(\omega)=\int dx^{0}\exp(-i\omega x^{0})$, made the identification $T=\int dx^{0}$ and used the explicit expression for the source $J({\bf x})=\lambda\delta^{3}({\bf x}-{\bf a})$.

All we need now is to compute the quantities ${\tilde{\cal G}}_{0}(0;{\bf b},{\bf a})$ and ${\tilde{\cal G}}_{0}(0;{\bf b},{\bf b})$. For this task we start by comparing the Fourier representation for $G_{0}(x,y)$,
%%%%%%%%%
\begin{eqnarray}
\label{zxc7a}
G_{0}(x,y)&=&-\int\frac{d\omega}{2\pi}\int\frac{d^{3}{\bf k}}{(2\pi)^{3}}\frac{e^{i{\bf k}({\bf x}-{\bf y})}}{\omega^{2}-{\bf k}^{2}-m^{2}}e^{-i\omega(x^{0}-y^{0})}\ ,
\end{eqnarray}
%%%%%%%%%
with the second equation (\ref{zxc3}), what leads to
%%%%%%%%%
\begin{eqnarray}
\label{zxc7b}
{\tilde{\cal G}}_{0}(\omega;{\bf x},{\bf y})&=&\int\frac{d^{3}{\bf k}}{(2\pi)^{3}}\frac{e^{i{\bf k}({\bf x}-{\bf y})}}{{\bf k}^{2}+(m^{2}-\omega^{2})}\ .
\end{eqnarray}
%%%%%%%%%

As discussed in reference \cite{BaroneHidalgo1}, for the case where ${\bf x}\not={\bf y}$, the integral in (\ref{zxc7b}) can be calculated by using dimensional regularization. The result for $\omega=0$, ${\bf x}={\bf b}$ and ${\bf y}={\bf a}$ (with ${\bf b}\not={\bf a}$) is,
%%%%%%%%%
\begin{eqnarray}
\label{zxc7c}
{\tilde{\cal G}}_{0}(0;{\bf b},{\bf a})&=&\frac{1}{4\pi}\frac{1}{|{\bf b}-{\bf a}|}\exp\!\Big(\!\!-m|{\bf b}-{\bf a}|\Big),\ \ {\bf b}\not={\bf a} .
\end{eqnarray}
%%%%%%%%%
For the case where ${\bf x}={\bf y}={\bf b}$, ${\tilde{\cal G}}_{0}(0;{\bf b},{\bf a})$ can be calculated with the analytic continuation of the integral (\ref{zxc7b}), as discussed in reference \cite{Kaku}. In the appendix we obtain the result
%%%%%%%%%
\begin{equation}
\label{zxc7d}
{\tilde{\cal G}}_{0}(0;{\bf x},{\bf x})=-\frac{m}{4\pi}.
\end{equation}
%%%%%%%%%

Defining $R=|{\bf a}-{\bf b}|$ and substituting (\ref{zxc7c}) and (\ref{zxc7d}) in (\ref{zxc6}) we get the interaction energy,
%%%%%%%%%
\begin{equation}
\label{asd2}
E_{int}(\mu,m,R)=\frac{\lambda^{2}}{32\pi^{2}}\frac{\mu}{1-[\mu m/(4\pi)]}\frac{1}{R^{2}}\exp{(-2mR)}\ ,
\end{equation}
%%%%%%%%%
which exhibits a faster decay with the distance $R$ in comparison with the Yukawa potential. Taking into account the restrictions mentioned in the sentence after Eq. (\ref{segundomodelo}), it is easy to show that the above interaction energy leads always to a non-divergent attractive force between the potential and the source.

Special cases are given by
%%%%%%%%%
\begin{eqnarray}
\label{asd3}
E_{int}(\mu,m=0,R)&=&\frac{\lambda^{2}\mu}{32\pi^{2}}\frac{1}{R^{2}}\cr\cr
E_{int}(\mu\to\infty,m,R)&=&-\frac{\lambda^{2}}{8\pi}\frac{\exp(-2mR)}{mR^{2}}\ .
\end{eqnarray}
%%%%%%%%%

The first result (\ref{asd3}) is a kind of attractive centrifugal potential (for $m=0$ we must have $\mu<0$) and corresponds to the case of a massless field. The second result is an attractive force and corresponds to the case of a perfect mirror. As discussed in reference \cite{Scardichio}, the massless case with perfect mirror leads to tachyonic field modes and should not be considered. This last case leads to a divergence in the energy (\ref{asd2}).

%%%%%%%%%
\section{Final Remarks}
\label{conclusion}
%%%%%%%%%
In conclusion, the computation of the modified scalar propagator due to the presence of a semi-transparent mirror was carried out. Its influence on the interaction energy between delta-like potentials and sources was investigated, and exact analytical results were found corresponding to the interacting potentials between classical objects in arbitrary dimensions. In what concerns the interaction between a $(D-1)-$dimensional potential and a point charge, the results turn out to be the generalization of the image method for a semi-transparent hyperplane mirror. The case of point-like potentials in $(3+1)$ dimensions gives rise to short range attractive interactions, with the massless case having the same spacial behavior as the one exhibited by the centrifugal potential (but with an attractive character).

Despite its theoretical interest, the use of these techniques to describe interactions intermediated by gauge bosons would be of great help in order to model measurable signatures of quantum effects like the Casimir effect between real conductors and graphene \cite{graphene}, among other things. Nevertheless, this kind of description still remain elusive because of the outstanding complexities presented by gauge fields. It is our hope that this work can also pave the way in this endeavor.

%%%%%%%%%
\begin{acknowledgments}
The authors would like to thank CAPES, FAPEMIG and CNPq for invaluable financial support and J.A. Helay\"el-Neto for valuable suggestions.
\end{acknowledgments}
%%%%%%%%%

%%%%%%%%%
%\appendix
%%%%%%%%%

\section*{Appendix}

In this appendix we compute the result (\ref{zxc7d}) by analytic continuation. For this task we first start with (\ref{zxc7b}), setting $\omega=0$ and ${\bf x}={\bf y}$,
%%%%%%%%%
\begin{equation}
{\tilde{\cal G}}_{0}(0;{\bf x},{\bf x})=\int\frac{d^{3}{\bf k}}{(2\pi)^{3}}\frac{1}{{\bf k}^{2}+m^{2}}\ .
\end{equation}
%%%%%%%%%

Using spherical coordinates, where $r=\sqrt{{\bf r}^{2}}$ is the radial coordinate, and integrating in the solid angle we have
%%%%%%%%%
\begin{equation}
\label{ap1}
{\tilde{\cal G}}_{0}(0;{\bf x},{\bf x})=\frac{1}{2\pi^{2}}\int_{0}^{\infty}dr\frac{r^{2}}{r^{2}+m^{2}}\ .
\end{equation}
%%%%%%%%%

Now we make use of the well known result (see for instance \cite{Kaku})
%%%%%%%%%
\begin{equation}
\label{ap2}
\int_{0}^{\infty}dr\frac{r^{\beta}}{(r^{2}+C^{2})^{\alpha}}=\frac{\Gamma\bigl((1+\beta)/2\bigr)\Gamma\bigl(\alpha-(1+\beta)/2\bigr)}{2(C^{2})^{\alpha-(1+\beta)/2}\Gamma(\alpha)}\ .
\end{equation}
%%%%%%%%%    

In expression (\ref{ap2}), the integral on the left hand side is well defined only for $2\alpha>1+\beta$ (on the contrary, the integrand diverges for large values of $r$). In spite of this, the right hand side is well defined for a wide range of values of $\alpha$ and $\beta$. So, in this sense, the right hand side of (\ref{ap2}) is the analytic extension of the integral in the left hand side. Taking $C=m$, $\alpha=1$ and $\beta=2$ in (\ref{ap2}), substituting the result in (\ref{ap1}), using the fact that $\Gamma(3/2)=\sqrt(\pi)/2$, $\Gamma(-1/2)=-2\sqrt(\pi)$ and $\Gamma(1)=1$ and taking into account the fact that $m>0$, we have the result (\ref{zxc7d}).

%%%%%%%%%

%%%%%%%%%

\end{document}